\documentclass[12pt]{article}
\def\email#1{\date{\tt#1}}
\def\address#1{\par\noindent#1\smallskip}

\begin{document}

\title{Bushes of normal modes - new dynamical objects in nonlinear mechanical
systems with discrete symmetry}
\author{George M.~Chechin, Alexandr V.~Gnezdilov, Vladimir P.~Sakhnenko \and
and Mikhail Yu.~Zekhtser}

\email{chechin@phys.rnd.runnet.ru}

\maketitle
\thispagestyle{empty}

\begin{abstract}
Bushes of normal modes represent exact mathematical objects
describing specific dynamical regimes in nonlinear mechanical
systems with point or space symmetry. In the present paper, we
outline the bush theory and illustrate it with some bushes of
small dimensions in octahedral mechanical structures.
\end{abstract}

\section{Introduction}

A new concept, ``bushes of normal modes," was introduced for
nonlinear mechanical systems with discrete symmetry in~\cite{Dan1,
Dan2}. A given bush represents a certain superposition of the
modes associated with different irreducible representations
(irreps) of the symmetry group~$G$ of the mechanical system in
equilibrium. The coefficients of this superposition are
time-dependent functions for which the exact ordinary differential
equations can be obtained. In this sense, the bush can be
considered as a dynamical object whose dimensionality is frequently
less than that of the original
mechanical system. The following propositions were justified in
previous papers~\cite{Dan1,Dan2,PhysD}:
\begin{enumerate}
\item A certain subgroup $G_D$ of the symmetry group $G$
corresponds to a given bush, and this bush can be excited by
imposing the appropriate initial conditions with the above symmetry
group $G_D \subset G$.
\item Each mode belonging to the bush possesses its own symmetry
group which is greater than or equal to the group $G_D$ of the whole
bush.
\item In spite of evolving mode amplitudes, the complete
collection of modes in the given bush is preserved in time and, in
this sense, the bush can be considered as a geometrical object.
\item The energy of the initial excitation is trapped in the bush,
i.e. it cannot spread to the modes which do not belong to the bush,
because of the symmetry restrictions.
\item As an indivisible nonlinear object, the bush exists because of
force interactions between the modes contained in it.
\item Taking into account the concrete type of interactions
between particles of the considered mechanical system can only
reduce the dimension of the given bush.
\item The extension of the bush can be realized as a result of the
loss of its stability which is accompanied by spontaneous breaking
of the bush symmetry (dynamical analog of phase transition).
\end{enumerate}

The special group-theoretical methods for finding bushes of modes
are discussed in~\cite{PhysD,IJNM}. The computer implementation of
these methods~\cite{Iso} (see also~\cite{Chechin}) allowed us to
find bushes of modes for wide classes of mechanical systems with
discrete symmetry. In particular, ``irreducible" bushes of
vibrational modes and symmetry determined similar non-linear
normal modes for all N-particle mechanical systems with the
symmetry of any of the 230 space groups were found in~\cite{IJNM}.
The bushes of vibrational modes of small dimensions were found and
classified into universality classes for all mechanical systems
with point groups of crystallographic symmetry in ~\cite{ENOC}.

\section{Outline of the bush theory}
\label{outline}

We consider  classical Hamiltonian systems of $N$ mass points
moving near the single equilibrium state which can be
characterized by a certain point or space symmetry group $G$. Let
the $3\times N$-dimensional vector,
\begin{equation}
\label{eq1} {\bf X}(t)=({\bf x}_1(t),{\bf x}_2(t),\ldots,{\bf
x}_N(t)),
\end{equation}
describe the displacements ${\bf x}_i(t)$ of all particles of our
mechanical system from their equilibrium positions. (Here we
denote by the three-dimensional vector ${\bf x}_i$ the displacement of
the $i$-th particle along the $X,Y$ and $Z$ axes).

The vector ${\bf X}(t)$ can be written as a superposition of all
basis vectors $ {\mbox{\boldmath$\varphi$}}_i^{(j)}$ of the
irreducible representations $\Gamma_j$ of the above mentioned
symmetry group $G$:
\begin{equation}
\label{eq2}{\bf
X}(t)=\sum_{j,i}\mu_i^{(j)}(t){\mbox{\boldmath$\varphi$}}_i^{(j)}.
\end{equation}
The coefficients $\mu_i^{(j)}(t)$ of this superposition depend on
time $t$, while the $3\times N$-dimensional time-independent
vectors ${\mbox{\boldmath$\varphi$}}_i^{(j)}$ are determined the
specific patterns of displacements of all particles of our
mechanical system.

Note that individual components of the basis vectors
${\mbox{\boldmath$\varphi$}}_i^{(j)}$ are often called
``symmetry-adapted coordinates". In particular, they can be normal
coordinates. Hereafter, the term, ``mode," means an arbitrary
superposition of basis vectors corresponding to a given irrep
$\Gamma_j$. As a result of this definition, every term
$\mu_i^{(j)}(t){\mbox{\boldmath$\varphi$}}_i^{(j)}$ in the right
hand side of Eq.(\ref{eq2}) is also a mode of the irrep
$\Gamma_j$. Sometimes, for brevity, we will refer to
$\mu_i^{(j)}(t)$ as a mode, but a reader must imagine
that this time-dependent coefficient is multiplied by the
appropriate $3\times N$-dimensional vector
${\mbox{\boldmath$\varphi$}}_i^{(j)}$ to give the mode in the
exact sense. We can also speak about {\it vibrational}
modes because only such type of symmetry-adapted (normal) modes
are considered in the present paper.

Every dynamical regime of the considered mechanical system can be
described by the appropriate time-dependent vector ${\bf X}(t)$
which determines a definite instantaneous  configuration of the
system. On the other hand, each instantaneous configuration
possesses a certain symmetry group $G_D$ (in particular, this
group may be trivial: $G_D=1$) which is a subgroup of the symmetry
group $G$ of the system in equilibrium ($G_D\subseteq G$).
Moreover, we can also ascribe a certain symmetry group to each
basis vector ${\mbox{\boldmath$\varphi$}}_i^{(j)}$ and to each
mode corresponding to a given irrep $\Gamma_j$ (remember that a mode
is a superposition of such vectors!), because the definite
instantaneous configurations correspond to them. The group $G_D$
contains all symmetry elements of group $G$ whose action does not
change this configuration.

Let us introduce, as it is usual in group theory, the
operators $\hat g$ associated with elements $g$ of group $G$
($g\in G$) which act on $3\times N$-dimensional vectors ${\bf
X}(t)$.\footnote{The symmetry elements $g\in G$ act on the vectors
of three-dimensional Euclidean space.} All elements $g\in G$ for
which
\begin{equation}
\label{eq3}\hat g{\bf X}(t)={\bf X}(t)
\end{equation}
form a certain subgroup $G_D\subseteq G$, and a
complete set of the above operators $\hat g$ ($\forall g\in G_D$)
represents the group $\widehat G_D$.

It can be shown that the symmetry group $G_D$ is preserved in time
in the sense that its elements {\it cannot disappear} during time
evolution. Actually, this property is a consequence of the
principle of determinism in classical mechanics.\footnote{The
phenomenon of spontaneous breaking of symmetry of a given
dynamical regime will be considered in the next section.} Thus,
the equation $\hat g{\bf X}(t)={\bf X}(t)$ $(g\in G_D)$, or,
formally,
\begin{equation}
\label{eq4} \widehat G_D{\bf X}(t)={\bf X}(t)
\end{equation}
is valid for every time $t$. As a consequence, we can classify the
different dynamical regimes in our nonlinear dynamical system,
described by the vectors ${\bf X}(t)$ from
Eqs.(\ref{eq1},\ref{eq2}), with the aid of {\it symmetry groups}
corresponding to them.

Using Eq.(\ref{eq4}), we can obtain the similar {\it invariance
conditions} for each individual irrep $\Gamma_j$ of the group $G$
(see details in~\cite{PhysD}):
\begin{equation}
\label{eq5} (\Gamma_j\downarrow
G_D){\mbox{\boldmath$\mu$}}_j={\mbox{\boldmath$\mu$}}_j.
\end{equation}
Here $(\Gamma_j\downarrow G_D)$ is a restriction of the irrep
$\Gamma_j$ to the subgroup $G_D$ of the group $G$, and
${\mbox{\boldmath$\mu$}}_j=(\mu_1^{(j)},\ldots,\mu_{n_j}^{(j)})$
is an {\it invariant vector} of $\Gamma_j$ ($n_j$ is the dimension of
this irrep).

To find all modes contributing to a given dynamical regime with
symmetry group $G_D$, i.e., for the vector ${\bf X}(t)$ from
Eq.(\ref{eq2}), we must solve linear algebraic
equations~(\ref{eq5}) for each irrep $\Gamma_j$ of the group $G$.
As a result of this procedure, the invariant vector
${\mbox{\boldmath$\mu$}}_j$ for some irreps $\Gamma_j$ can turn out to
be equal to zero. Such irreps do not contribute to the considered
dynamical regime. On the other hand, some nonzero invariant
vectors ${\mbox{\boldmath$\mu$}}_j$ for multidimensional irreps
may be of a very specific form because of definite relations between
their components (for example, certain components can be equal to
each other, or differ only by sign).

Actually, Eq.(\ref{eq5}) can be considered as a source
of certain selection rules for spreading excitation from the root
mode to a number of other (secondary) modes. Indeed, if a certain
mode with the symmetry group  $G_D$ is excited at the initial
instant (we call it the ``root" mode), this group determines the
symmetry of the whole bush. The condition that the appropriate
dynamical regime ${\bf X}(t)$ must be invariant under the action
of the above group $G_D$ leads to Eq.(\ref{eq4})
and then to Eq.(\ref{eq5}). If the vector
${\mbox{\boldmath$\mu$}}_j$ for a given irrep $\Gamma_j$ proves to
be a zero vector, then there are no modes belonging to this irrep
which contribute to ${\bf X}(t)$, i.e., the initial excitation {\it
cannot spread} from the root mode to the secondary modes
associated with the irrep $\Gamma_j$.

Note that basis vectors associated with a given irrep $\Gamma_j$
in Eq.(\ref{eq2}) turn out to be equal to zero when this irrep is not
contained in the decomposition of the full vibrational irrep
$\Gamma$ into its irreducible parts $\Gamma_j$. This is a source
of the additional selection rules which reduce  the  number of
possible vibrational modes in the considered bush. Trying every
irrep $\Gamma_j$ in Eq.(\ref{eq5}) and analyzing the above
mentioned decomposition of the vibrational representation
$\Gamma$, we obtain the whole bush of modes with the symmetry
group $G_D$  in the explicit form.

Let us return to Eq.(\ref{eq2}). We speak about geometrical
aspects of the bush theory when concentrate our attention on basis
vectors ${\mbox{\boldmath$\varphi$}}_i^{(j)}$, and we speak about
dynamical aspects of this theory when we focus on time-dependent
coefficients $\mu_i^{(j)}(t)$, which will be also called ``modes".

If interactions between the particles of our mechanical system are
known, {\it exact} dynamical equations describing the time
evolution of a given bush can be written.

Two types of interactions between modes in nonlinear Hamiltonian
system are discussed in~\cite{PhysD}, namely, {\it force}
interactions and {\it parametric} interactions. We can illustrate
the difference between these types of modal interactions using a
simple example.

Let us consider two different linear oscillators whose coupling is
described by only one anharmonic term, $U=-\gamma \mu_1^2\mu_2$, in
the potential energy. Dynamical equations for this system can
be written as follows:
\begin{eqnarray}
\label{eq6a} \ddot\mu_1+\omega_1^2\mu_1&=&2\gamma\mu_1\mu_2, \\
\label{eq6b} \ddot\mu_2+\omega_2^2\mu_2&=&\gamma\mu_1^2.
\end{eqnarray}
Here $\gamma$ is an arbitrary constant characterizing the strength
of the interaction of the oscillators. We can suppose that
Eqs.(\ref{eq6a},\ref{eq6b}) describe the dynamics of two
modes $\mu_1(t)$ and $\mu_2(t)$ in a certain mechanical system.

An essential disparity between modes $\mu_1(t)$ and $\mu_2(t)$ can
be seen from the above equations. Indeed, if we excite the mode
$\mu_1(t)$ at the initial instant ($\mu_1(t_0)\ne 0$), the mode
$\mu_2(t)$ cannot be equal to zero (even if it was zero at
$t=t_0$!) because a {\it nonzero force} $-\frac{\partial
U}{\partial\mu_2}=\gamma\mu_1^2$ appears in the right hand side of
Eq.(\ref{eq6b}) since $\mu_1(t)\not\equiv 0$. In other words, the
dynamical regime $\mu_1(t)\not\equiv 0, \mu_2(t)\equiv 0$ cannot
exist because of the contradiction with Eq.(\ref{eq6b}). Unlike
this, the dynamical regime $\mu_2(t)\not\equiv 0, \mu_1(t)\equiv
0$ can exist because such a condition does not contradict
equations~(\ref{eq6a}) and~(\ref{eq6b}). We can say that now there
is no force in the right hand side of Eq.(\ref{eq6b}) because
$\frac{\partial U}{\partial\mu_2}\equiv 0$ as a consequence of the
identity $\mu_1(t)\equiv 0$.

In the last case, the dynamical regime of the  system
(\ref{eq6a},\ref{eq6b}) represents a harmonic oscillation only of
the second variable:
\begin{equation}
\label{eq7}\mu_2(t)=A\cos(\omega_2t+\delta),
\end{equation}
where $A$ and $\delta$ are two arbitrary constants.

Thus, there is {\it force} action from the mode $\mu_1(t)$ on the
mode $\mu_2(t)$, but not vice versa. We proved in~\cite{PhysD}
that such a situation can be realized only in the case where the symmetry
group of the mode $\mu_1(t)$ is less than or equal to that of the
mode $\mu_2(t)$.\footnote{Remember, that speaking about a mode we
must take into account that our time-dependent coefficient is
multiplied by the appropriate basis vector of a certain  irrep of
the group $G$, and namely this vector determines the symmetry
group of the considered mode.}

Nevertheless, the mode $\mu_2(t)$ can excite the mode $\mu_1(t)$
under certain circumstances. Indeed, substituting the
solution~(\ref{eq7}) of Eq.(\ref{eq6b}) into  Eq.(\ref{eq6a}), we
obtain
\begin{equation}
\label{eq8}\ddot\mu_1(t)+\left[\omega_1^2-2A\gamma\cos(\omega_2t+\delta)\right]\mu_1=0.
\end{equation}
By means of simple algebraic transformations this equation can be
converted to the Mathieu equation in its standard form:
\begin{equation}
\label{eq9}z''+\left[a-2q\cos(2\tau)\right]z=0,
\end{equation}
where $z=z(\tau)$. But in the ($a-q$) plane of the Mathieu
equation~(\ref{eq9}) there exist domains of stable and unstable
movement. If pertinent parameters ($\omega_1, \omega_2, \gamma$)
of our dynamical system~(\ref{eq6a},\ref{eq6b}) have values such
that corresponding parameters $a$ and $q$ of
Eq.(\ref{eq9}) get into an unstable domain, then the nonzero
function $z(\tau)$ and, therefore, the mode $\mu_1(t)$ appears. In
other words, the initial dynamical regime  $\mu_1(t)\equiv 0,
\mu_2(t)\not\equiv 0$ loses its stability, and a new dynamical
regime $\mu_1(t)\not\equiv 0, \mu_2(t)\not\equiv 0$ arise {\it
spontaneously} for definite values of the parameters of
Eqs.(\ref{eq6a},\ref{eq6b}). Since this phenomenon is similar to
the well-known parametric resonance, we can speak, in such a case,
about {\it parametric action} from the mode $\mu_2(t)$ on the mode
$\mu_1(t)$.

The characteristic property of the parametric interaction is that
the appropriate force ($-\frac{\partial U}{\partial
\mu_1}=2\gamma\mu_1\mu_2$, in our case) vanishes when the mode
($\mu_1$, in our case), on which this force acts, becomes zero.
The following important result was proved in~\cite{PhysD}: the
mode of lower symmetry acts on the mode of higher symmetry
by {\it force} interaction, while the mode of higher symmetry can
act on the mode of lower symmetry only {\it parametrically}.
Consequently, if the parametric excitation of a certain mode does
take place, this phenomenon must be by necessity be accompanied by
{\it spontaneous breaking} of symmetry of the mechanical system
vibrational state.

Thus, the initially excited dynamical regime can lose its
stability because of parametric interactions with some zero modes
and, as a result, can transform spontaneously into another
dynamical regime, described by a greater number of
dynamical variables, with appropriate lowering of symmetry.
Obviously, we may treat such phenomenon as a dynamical analog of
a phase transition.

\section{Examples of bushes of vibrational modes}

We consider a mechanical system  of six mass points (particles)
whose interactions are described by a pair isotropic potential
$u(r)$ where $r$ is the distance between two particles. We suppose
that in the {\it equilibrium state} these particles form a regular
octahedron with edge $a_0$ which can be imagined in the
following way. Let us introduce a Cartesian coordinate system.
Four particles of the above octahedron lie in the $XY$ plane and form
a square with edge $a_0$. Two other particles lie on $Z$ axis
and we will speak about the ``top particle" and the ``bottom
particle" with respect to the direction  of the $Z$ axis. Obviously,
the distance between each of these two particles and any of the
four particles in the $XY$ plane is equal to $a_0$.

The point symmetry group of the octahedral equilibrium
configuration is $O_h$ and all bushes of vibrational modes in the
considered system are described by the certain subgroups $G_D$ of
this parent group. In the present paper, we consider only three
bushes\footnote{We write the symmetry group $G_D$ of the bush in
square brackets next to its symbol.} $B1[O_h]$, $B2[D_{4h}]$ and
$B4[C_{4v}]$ (the complete list includes 18 different by symmetry
bushes of vibrational modes whose dimensions vary from 1 to
12). The symmetry groups of the above bushes are connected with
each other by the following group-subgroup relation:
$C_{4v}\subset D_{4h}\subset O_h$.

The geometrical forms of our mechanical system in the {\it
vibrational state}, corresponding to these bushes, can be revealed
from the appropriate symmetry groups $G_D$.

The one-dimensional bush $B1[O_h]$ consists of only one
(``breathing") mode. This nonlinear dynamical regime ${\bf X}(t)=
\mu_1^{(1)}(t){\mbox{\boldmath$\varphi$}}_1^{(1)}$ describes
evolution of a regular octahedron whose edges $a=a(t)$ periodically
change in time.

The two-dimensional bush $B2[D_{4h}]$ describes a dynamical regime
with two degrees of freedom: ${\bf X}(t)=
\mu_1^{(1)}(t){\mbox{\boldmath$\varphi$}}_1^{(1)}+\mu_1^{(5)}(t){\mbox{\boldmath$\varphi$}}_1^{(5)}$.
The symmetry group $G_D=D_{4h}$ of this bush contains the 4-fold
axis coinciding with the $Z$ coordinate axis and the mirror plane
coinciding with the $XY$ plane. This symmetry group restricts
essentially the form of the polyhedron describing our mechanical
system in the vibrational state. Indeed, the presence of the 4-fold
axis demands that the quadrangle in the $XY$ plane be a square. Because
of the same reason, the four edges connecting the particles in the $XY$
plane (vertices of the above square) with the top particle lying
on $Z$ axis must be of the same length which we denote by $b(t)$.

Similarly, let the length of the edges connecting the bottom
particle on the $Z$ axis with any of the 4 particles in the $XY$ plane be
denoted by $c(t)$. In our present case of the bush $B2[D_{4h}]$,
$b(t)=c(t)$ for any time $t$ because of the presence of the
horizontal mirror plane in the group $G_D=D_{4h}$. But for the
three-dimensional bush $B4[C_{4v}]$, described by ${\bf X}(t)=
\mu_1^{(1)}(t){\mbox{\boldmath$\varphi$}}_1^{(1)}
+\mu_1^{(5)}(t){\mbox{\boldmath$\varphi$}}_1^{(5)}
+\mu_3^{(10)}(t){\mbox{\boldmath$\varphi$}}_3^{(10)}$, this mirror
plane is absent and, therefore, $b(t)\ne c(t)$.

Let us also introduce two heights, $h_1(t)$ and $h_2(t)$,
corresponding to the perpendiculars dropped, respectively, from the top
and bottom vertices of our polyhedron in the $XY$ plane. Now we
can write the dynamical equations of the above bushes in terms of
pure geometrical variables $a(t), b(t), c(t), h_1(t)$ and
$h_2(t)$.

We choose $a(t)$ and $h(t)\equiv h_1(t)\equiv h_2(t)$ as dynamical
variables for describing the two-dimensional bush $B2[D_{4h}]$ and
$a(t), h_1(t)$ and $h_2(t)$ as dynamical variables for describing
the three-dimensional bush $B4[C_{4v}]$. Using these variables we
can write down the potential energy for our bushes of vibrational
modes as follows: $$
\begin{array}{lcl}
B1[O_h]&:& V_{B1}(a)=12u(a)+3u(\sqrt{2}\,a),\\
B2[D_{4h}]&:&V_{B2}(a,h)=4u(a)+2u(\sqrt{2}\,a)+8u\left(\sqrt{h^2+\frac{a^2}{2}}\,\right)
+u(2h),\\
B4[C_{4v}]&:&V_{B4}(a,h_1,h_2)=4u(a)+2u(\sqrt{2}\,a)+4u(b)+4u(c)+u(h_1+h_2),
\end{array}
$$

where
$b=\sqrt{\frac{a^2}{2}+\left(\frac{5}{4}h_1-\frac{1}{4}h_2\right)^2}$,
$c=\sqrt{\frac{a^2}{2}+\left(\frac{5}{4}h_2-\frac{1}{4}h_1\right)^2}$.

Then with the aid of the Lagrange method, we can obtain the
following dynamical equations for the above bushes of vibrational
modes:
\begin{equation} \label{ur}
\begin{array}{ll}
B1[O_h]:&\\ \qquad \ddot a= & -4u'(a)-\sqrt{2}u'(\sqrt{2}a); \\
B2[D_{4h}]:& \\ \qquad \ddot a= &
-2u'(a)-\sqrt{2}u'(\sqrt{2}a)-2u'(b)\frac{a}{b}, \\ \qquad \ddot
h= & -4u'(b)\frac{h}{b}-u'(2h); \\ B4[C_{4v}]:& \\ \qquad \ddot a=
& -2u'(a)-\sqrt{2}u'(\sqrt{2}a)-u'(b)\frac{a}{b}-u'(c)\frac{a}{c},
\\ \qquad \ddot h_1= & -u'(b)\frac{5h_1-h_2}{b}-u'(h_1+h_2), \\
 \qquad \ddot h_2= & -u'(c)\frac{5h_2-h_1}{c}-u'(h_1+h_2).
\end{array}
\end{equation}

Thus, we obtain the dynamical equations of our bushes of
vibrational modes in terms of variables with explicit geometrical
sense. Each bush describes a certain nonlinear dynamical regime
corresponding to such a vibrational state of the considered mechanical
system, that at any fixed time the configuration of this system is
represented by a definite polyhedron with symmetry group $G_D$ of
a given bush.

We can write dynamical equations for the above bushes in terms of
vibrational modes as well. In spite of the more complicated form,
these equations turn out to be more useful for the bush
theory, since they allow us to decompose the appropriate
nonlinear dynamical  regimes into modes of different importance
for the case of small oscillations -- root modes and
secondary modes of different orders~\cite{PhysD}. As an example,
we write below the dynamical equations for the bush $B4[C_{4v}]$
in terms of its three modes, $\mu_3^{(10)}(t)\equiv \gamma(t)$
(root mode), $\mu_1^{(1)}(t)\equiv\mu(t)$,
$\mu_1^{(5)}(t)\equiv\nu(t)$ (secondary modes):
$$
\begin{array}{ll}
\ddot{\mu}=&-\frac{1}{6}(4\sqrt{2}u'(a)+4u'(\sqrt{2}a)+2u'(h_1+h_2)+
\frac{u'(b)}{b}(2\sqrt{2}a+5h_1-h_2)+\\& \\ \qquad&
\frac{u'(c)}{c}(2\sqrt{2}a+5h_2-h_1)),\\&\\

\ddot{\nu}=&-\frac{1}{6}(2\sqrt{2}u'(a)+2u'(\sqrt{2}a)-2u'(h_1+h_2)+
\frac{u'(b)}{b}(\sqrt{2}a-5h_1+h_2)+\\& \\ \qquad&
\frac{u'(c)}{c}(\sqrt{2}a-5h_2+h_1)),\\&\\

\ddot{\gamma}=&\frac{1}{4}\left(u'(b)\frac{5h_1-h_2}{b}-u'(c)\frac{5h_2-h_1}{c}\right).
\end{array}
$$
Here $$
\begin{array}{c}
\label{201} a=\sqrt{2}(r_0+\mu+\nu);
b=\sqrt{(r_0+\mu+\nu)^2+(r_0+\mu-2\nu-3\gamma)^2};\\
c=\sqrt{(r_0+\mu+\nu)^2+(r_0+\mu-2\nu+3\gamma)^2};\\
h_1=r_0+\mu-2\nu-2\gamma; h_2=r_0+\mu-2\nu+2\gamma;
r_0=\frac{a_0}{\sqrt{2}}.
\end{array}
$$

In the previous section, the problem of existence of bushes of
vibrational modes was studied by group-theoretical methods
only. In contrast to this, examination of the stability
of the bushes depends essentially on the concrete type of
interactions in the considered system, and we suppose that they
can be described by a Lennard-Jones potential. Then the bush
stability will be analyzed with the aid of numerical methods.

Let us excite a given bush $B[G_D]$ with the aid of the following
{\it initial condition}. Note that the root mode of each of the
above considered bushes is determined by a {\it single}
time-dependent coefficient $\mu_i^{(j)}(t)$ whose initial value at
$t=t_0$ we will denote by the symbol $\mu_0$ $(\mu_0 \equiv
\mu_i^{(j)}(t_0))$. At the initial instant $t=t_0$, we fix
the coordinates of all particles of the mechanical system in such a
way that their {\it displacements} correspond to the appropriate
root mode with amplitude $\mu_0$, while their velocities are equal
to {\it zero}. Namely this choice of initial conditions determines
the {\it way} of excitation of a given bush.

Using these initial conditions we solve numerically the exact
dynamical equations of the considered mechanical system with 18
degrees of freedom, and analyze the set of nonzero modes
$\mu_i^{(j)}(t)$ in the decomposition (\ref{eq2}) of the vector
${\bf X}(t)$ obtained as a result of this solving. Then we
gradually increase the value $\mu_0$ and repeat the procedure
just described until the number of nonzero modes  $\mu_i^{(j)}(t)$, at
some value $\mu_0=R$, becomes larger than that of the bush
$B[G_D]$. We will refer to $R$ as the {\it threshold} of stability
of the given bush. Obviously, in such a way we obtain the upper
boundary of the first stability region of $B[G_D]$ in
one-dimensional space of all possible values $\mu_0 \; (0< \mu_0
\leq R)$.\footnote{Note that we do not study the {\it other}
possible regions of stability of the given bush $B[G_D]$ in the
present paper.}

For $\mu_0>R$ a new bush $\widetilde{B}$[$\widetilde{G_D}$],
including the old bush $B[G_D]$, appears and its  symmetry
$\widetilde{G_D}$ is {\it lower} than that of this old bush
($\widetilde{G_D}\subset G_D$), because all modes with symmetry
higher than or equal to $G_D$ are already contained in $B[G_D]$.

Thus, the loss of stability of a given bush is accompanied by the
spontaneous breaking of symmetry of the initially excited
dynamical regime, described by this bush. We already discussed
this phenomenon  in Sec.\ref{outline} and concluded
that its cause is analogous to that of this parametric resonance.

We can also say this in other words. A given bush $B[G_D]$
represents a certain dynamical regime in the considered mechanical
system. Its modes interact with other modes which do not belong to
$B[G_D]$, but these interactions must be of parametric (not
force!) type only. For the appropriate initial conditions we can
get into a region of unstable movement. As a result, some new
modes are excited which were forbidden by principle of determinism
of classical mechanics. Then we can speak about the loss of
stability of the original bush $B[G_D]$ and its transformation
into a larger bush $\tilde{B}[\widetilde{G_D}]$ with
$(\widetilde{G_D} \subset G_D)$.

The Lennard-Jones potential can be written in the form,
\begin{equation}
\label{eq10p} u(r)=\frac{A}{r^{12}}-\frac{B}{r^6},
\end{equation}
 with $A=B=1$ (because of the appropriate scaling transformation of
space and time variables). Our numerical experiments give the
following values of the threshold $R$ for the above considered
bushes:
\begin{equation}
\label{102} R[B1]=0.001,\qquad R[B2]=0.009,\qquad R[B4]=0.003.
\end{equation}

Note that these values of $R$ are finite but very small in
comparison with the edge $a_0=1.117$ of our octahedral structure
in equilibrium. Displacements of particles corresponding to the
thresholds~(\ref{102}) can be obtained by dividing these three
values of $R$ by the numbers $\sqrt{6}$, $\sqrt{12}$, $\sqrt{12}$,
respectively. We want to note in this connection, that all
octahedral molecules, known to us at the present time, possess an
atom in the center of the octahedron. This fact suggests that the
stability of such structures can be greater than
that of the mechanical system considered up to this point. Taking
into account this hint, we examined bush stability for the
mechanical structure with the particle in the center of the
octahedron, supposing that this additional (seventh) particle is
described by the Lennard-Jones potential different from that for
six peripheral particles. Namely, we assume that $A=1$, but $B>1$
in Eq.(\ref{eq10p}). Such an assumption provides us a possibility of
making the attractive part of the potential of the centered particle
greater than that of peripheral atoms in spite of the same
repulsive part.

For such a centered structure the threshold values $R$ can be
essentially greater than those for the structure without the particle
in the center of the octahedron. Indeed, we obtained that
$R[B1]=1.010$, $R[B2]=0.011$, $R[B4]=0.118$ for $B=5.5$. Note that
the dependence of the threshold $R$ on the value $B$ is
essentially different for different bushes and can be
nonmonotonic.

\bigskip

\address{ George M.~Chechin, Rostov State University, Rostov-on-Don, Russia}
\address{ Alexandr V.~Gnezdilov, Rostov State University, Rostov-on-Don, Russia}
\address{ Vladimir P.~Sakhnenko, Institute of Physics of Rostov State University, Rostov-on-Don, Russia}
\address{ Mikhail Yu.~Zekhtser, Institute of Physics of Rostov State University, Rostov-on-Don, Russia}


\begin{thebibliography}{15}
\bibitem{Dan1}
V.P. Sakhnenko, G.M. Chechin, Symmetrical selection rules in
nonlinear dynamics of atomic systems. {\it Dokl. Akad. Nauk} {\bf
330}, 308 (1993). [Phys. Dokl. {\bf 38}, 219 (1993)].
\bibitem{Dan2}
V.P. Sakhnenko, G.M. Chechin, Bushes of modes and normal modes for
nonlinear dynamical systems with discrete symmetry. {\it Dokl.
Akad. Nauk}. {\bf 338}, 42 (1994). [Phys. Dokl. {\bf 39}, 625
(1994)].
\bibitem{PhysD}
G.M. Chechin, V.P. Sakhnenko, Interactions between normal modes in
nonlinear dynamical systems with discrete symmetry. Exact results.
{\it Physica~D.} {\bf 117}, 43 (1998).
\bibitem{IJNM}
G.M. Chechin, V.P. Sakhnenko, H.T. Stokes, A.D. Smith, D.M. Hatch,
Non-linear normal modes for systems with discrete symmetry. {\it
Int.~J.~Non-Linear~Mech}. {\bf 35}, 497 (2000).
\bibitem{Iso}
The software package, {\it ISOTROPY}, is available on the Internet
at http://www.physics.byu.edu/$\sim\,$stokesh/isotropy.html~.
\bibitem{Chechin}
G.M.~Chechin, Computers and group-theoretical methods for studying
structural phase transition. {\it Comput. Math. Applic.} {\bf 17},
255 (1989).
\bibitem{ENOC}
G.M.~Chechin, V.P.~Sakhnenko, M.Yu.~Zekhtser, H.T.~Stokes,
S.~Carter, D.M.~Hatch, Bushes of normal modes for nonlinear
mechanical systems with discrete symmetry. {\it World Wide Web
Proceedings of the 3rd ENOC conference} http://www.midit.dtu.dk
\end{thebibliography}
\end{document}